# Towards a definition of climate science

## Valerio Lucarini*

Department of Earth, Atmospheric and Planetary Sciences, Building
54 Room 17-19, Massachusetts Institute of Technology, 02138
Cambridge, Massachusetts, USA
e-mail: lucarini@mit.edu

**Abstract:** The intrinsic difficulties in building realistic climate models and in providing complete, reliable and meaningful observational datasets, and the conceptual impossibility of testing theories against data imply that the usual Galilean scientific validation criteria do not apply to climate science. The different epistemology pertaining to climate science implies that its answers cannot be singular and deterministic; they must be plural and stated in probabilistic terms. Therefore, in order to extract meaningful estimates of future climate change from a model, it is necessary to explore the model's uncertainties. In terms of societal impacts of scientific knowledge, it is necessary to accept that any political choice in a matter involving complex systems is made under unavoidable conditions of uncertainty. Nevertheless, detailed probabilistic results in science can provide a baseline for a sensible process of decision making.

## 1 Introduction

> "In order to protect the environment, the precautionary principle approach shall be widely applied by States according to their capabilities. Where there are threats of serious or irreversible damage, lack of full scientific certainty shall not be used as a reason for postponing cost-effective measures to prevent environmental degradation."
>
> Principle 15, United Nations Conference on Environment and Development (Rio de Janeiro, 1992).

The climate is defined as the mean physical state of the climatic system, which is constituted by atmosphere, hydrosphere, cryosphere, lithosphere and biosphere, which are intimately interconnected. Therefore, the climate is determined by a set of time-averages of quantities that describe the structure and the behaviour of the various parts of the climatic system, as well as by the correlations among them (Peixoto and Oort, 1992).

---

* Present address: Via Palestro 7, 50123 Firenze, Italy.

In the very definition of climate there is an ambiguity and an element of subjectivity because the extension of the time interval over which the statistics are made is not determined a priori, but is operationally chosen depending on the goal of the research.

The presence of such a weak foundation strongly determines all the features of climate science. This weakness does not imply that this is a bad science, as stated by various politicians and opinion makers around the world, but it is a natural consequence of the fact that the system, subject to the studies of the climate scientists, is extremely complex. This complexity of the climatic system is such that the feedback between the various parts play an essential role. It makes little sense to define single elements and processes, when it is more sensible to consider it as a non separable 'body', a 'living organism' which cannot be solved, i.e. explained in simple terms, as the origin of the word 'complex'[1] explains. Therefore, it is conceptually incorrect to expect that climate science could provide answers having comparable precision and similar structure to those provided by sciences that investigate less complex systems.[2] The complete understanding of the climate system is an open problem that may never be solved: this makes the study of the climate a scientific enterprise of exceptional interest. The urgency and the pressure for providing the policymakers with information that is necessary for the implementation of long-sighted policies gives climate science an extremely relevant sociopolitical role which, in turn, contributes to shaping the structure, goals and priorities of climate science. The latest report (Houghton *et al*., 2001) of the International Panel on Climate Change (IPCC; see, http://www.ipcc.ch) has recently been criticized for lacking quantitative estimations of the uncertainties of the projections of climate change (Reilly *et al*., 2001). An assessment of such uncertainties is needed in order to provide the governments with clear baselines to be able to initiate a process of rational decision making.

In order to correctly interpret the subsequently presented results obtained by some research teams in terms of quantitative evaluation of the uncertainties relating to the increase in mean global temperature a century from now, it is necessary to first present the main features of climate science.

## 2   Uncertainties in climate science

Due to the complexity of the system, climate dynamics is chaotic and is characterized by a large natural variability on different temporal scales that would cause non-trivial difficulties in detecting trends in statistically relevant terms, even if the observational data were absolutely precise.

The actual situation is much more problematic because even for the atmosphere, which is the observationally best-known component of the climatic system, the database of observations having global extension, good reliance and good temporal frequency go back in time no more than 4–5 decades. For each location, these observations are essentially temporally-averaged data for temperature, pressure and precipitation, which have been collected by several research centres, especially meteorological institutes, around the globe.

These collections of data about past climatology usually feature a relatively low degree of reciprocal synchronic coherence and individually present problems of diachronic coherence, due to changes in the strategies of data gathering with time. Furthermore, especially for the oldest years, some data are not available: they have been lost or their reliability is very low due to past technical failures or simply because they have not been collected.

The most precise instruments for the detailed simulation of climate dynamics are the General Circulation Models (GCMs), which describe the coupled evolution of the various components of the climatic system through the inclusion of a mathematical description of the main physical, chemical and biological processes. The complexity of these models is such that they generate a natural variability which is comparable to the observed one.

The quantification of many processes playing major roles in climate dynamics is still clearly incomplete. In term of chemistry and biology, there is great uncertainty about the determination of the natural emissions of greenhouse gases and the absence of a good, relatively detailed understanding of the carbon cycle. There are also substantial uncertainties on the correct values to assign to fundamental parameters of the physics of the system, like climate sensitivity[3] and the efficiency of the oceanic heat uptake from the atmosphere. It is not clear how the dynamics of these processes depend on climatic variables, so that it is not easy to understand how intense and of which sign —positive or negative — could be the feedbacks they could trigger in the process of climatic change.

It is also not possible to rule out the fact that essential elements may have not yet been discovered, such as nonlinearities able to generate so-called climatic surprises, i.e. rapid climatic changes that can take place in conditions of increased instability. A well-known example of these phenomena is the breakdown of the thermohaline circulation, whose occurrence could drive Northern Europe to a much colder climate than at present (Rahmstorf, 1997).

The limitations of the present GCMs are not just due to the previous conceptual difficulties: many of the known processes are implemented in the models with simplified dynamics in order to reduce the computer time needed for each simulation.

Each study, aimed at providing possible future scenarios, should also take into consideration the uncertainties related to future anthropogenic emissions of greenhouse gases. Since these generate the forcing which drives the system out of its natural equilibrium, it is reasonable to expect that the uncertainties in anthropogenic emissions have a major role in determining the uncertainty in climatic response. Consistent with the precision we require in describing processes of analogous importance (a sort of respect of the degree of complexity could be invoked), the data fed into the model to describe the anthropogenic emissions, considered in this context as the results of the process named 'human society', should be generated by a global economic model. Therefore, the uncertainties in the emissions should result from the uncertainties intrinsic to the economic model. The 'organicity' of the climate system requires all parts to be interconnected, so that the economic model should be coupled to the model describing the strictly speaking natural phenomena; for this purpose a good evaluation of the costs of the impacts of climate change is of the utmost importance.

Considering greenhouse gas emission only as an uncoupled factor capable of influencing the climate is conceptually unsatisfying and dangerous, since it implicitly assumes that human choices could be independent of the state of the environment where they themselves live.

In first approximation, the $CO_2$ emissions are closely related to global economic quantities, such as economic growth, energy-intensiveness of the economy, and shares of the various sources of energy.

The emissions of other greenhouse gases, such as $CH_4$, CFCs and $N_2O$, which now already contribute to a good part of the total positive radiative forcing, as well as aerosol emissions,[4] depend more precisely on the details of the economy.

The presence of structural uncertainties (due to the choices made when a model is built on which processes and feedbacks are described and how they are described) and of parametric uncertainties (due to the lack of knowledge on quantities which characterize the climatic system), implies that every model used to generate projections about future climate change is a priori false, or better, weak in its descriptive power. Climate science does not have a laboratory where theories could be tested against experiments; every model can be tested only against data from the past, which are not necessarily precise. The natural variability of both the model and of the real system contributes to blur the line between a failed and a passed test. Anyway, a positive result would not at all guarantee that the model is able to provide good future projections while at most we can conclude from a negative result that the model does not work properly.

The distance from Galilean science is so wide that it is impossible to apply the usual scientific validation criteria to the results of climate science.

The different epistemology pertaining to climate science implies that its answers cannot be singular and deterministic, while they must be plural and stated in probabilistic terms.

The necessity of providing explicit suggestions for the formulation and the implementation of global long-sighted policies has been a fundamental stimulation to climate science. This original aspect of contamination is clearly identifiable, observing the strong interaction and close interconnection between the scientific community, public opinion, governments, and private companies. This linkage touches even what the scientists have historically been very jealous of: determination of interesting problems and the creation of conceptual instruments used in defining the level of precision of the results that is required in research, and of conceptual instruments used in determining the level of precision that has been obtained in the single project.

## 3  A possible strategy

The aforementioned discussion explains why it is not sensible to expect to obtain qualitatively better results with the availability of more and more powerful computers alone, which would make it possible to get a finer resolution of the 3-D grid describing the planet and to have a more precise implementation of the single processes in the models.

While there are instruments to analyse the uncertainties arising from the natural variability, it is more difficult to deal with the structural and parametric uncertainties of the models. These uncertainties are intrinsic and thinking of them as obstacles preventing us from reaching some form of truth is epistemologically incorrect since this approach reflects a reductionistic attitude which, as already explained, is in this context quite misleading.

An epistemologically correct analysis takes into account those uncertainties and tries to figure out how they influence the uncertainties on climatic change projections: the scientifically-relevant and sensitive results can be expressed only in probabilistic terms.

The structural uncertainties cannot be studied using only one model: one can expect to analyse them by comparing different models following a horizontal and vertical conceptual hierarchical path. The horizontal comparison is the comparative study of the results generated by models sharing a roughly common level of complexity, but having been implemented in different ways by different people. The vertical comparison is the comparative study of the results obtained by a family of models, each built as an extension and complexification of another one starting from an initial simple parent,[5] thus creating a natural hierarchy of increasing complexity.

The analysis of parametric uncertainties is conceptually simpler and can be thought of as a study where several runs of the same model evolve from the same initial conditions but with different values of the most relevant uncertain parameters. The uncertainty on a parameter becomes a probability density function related to its value; if independent studies restrict the value of the parameter within a certain interval (Forest *et al.*, 2001; 2002), the above-mentioned function should reduce to zero outside such an interval. By using a Monte Carlo method,[6] it is possible to compile the statistics for the most relevant projected quantities, for which the probability distribution functions are obtained. Setting all the parameters at their mean value except one, it is possible to understand how strongly the uncertainty of the latter influences the uncertainties on the projections.

These experiments of experiments need many runs in order to provide statistically relevant results. Therefore, single research institutes do not have enough computer power to study parametric uncertainties on the best GCMs available, since they need time periods of the order of months just to complete one simulation even on powerful machines. On the contrary, simpler models gauged to give results compatible with those coming from hierarchically higher relatives can be used.

Two studies where parametric uncertainties have been treated in the aforementioned way have been recently presented. The parameters' values have been confined within intervals deduced from either independent scientific investigations or, where this was not possible, from the judgment of experts in the field (Morgan and Henrion, 1990).

In Wigley and Raper (2001), a 2-D climate model (Wigley and Raper, 1992) has been used to produce a probability density function of the mean global temperature change between 1990 and 2100. The authors have considered uncertainties affecting the climate sensitivity, the carbon cycle, the oceanic heat uptake from the atmosphere, the net effects of the aerosols, and the future paths of emissions of greenhouse gases. In Figure 1 we present the results of this study for the distributions of the 1990–2100 warming for three cases, which differ on the choices of the probability distribution of value of a parameter and on the inclusion of the analysis if the effect of the presence of uncertainty of another one (for more details, see, Wigley and Raper, 2001). In the most complete and realistic case (represented in Figure 1 by the thick line), the 90% confidence interval of the 1990–2100 mean temperature increase is 1.7–4.9 °C, while the median value is 3.1 °C.

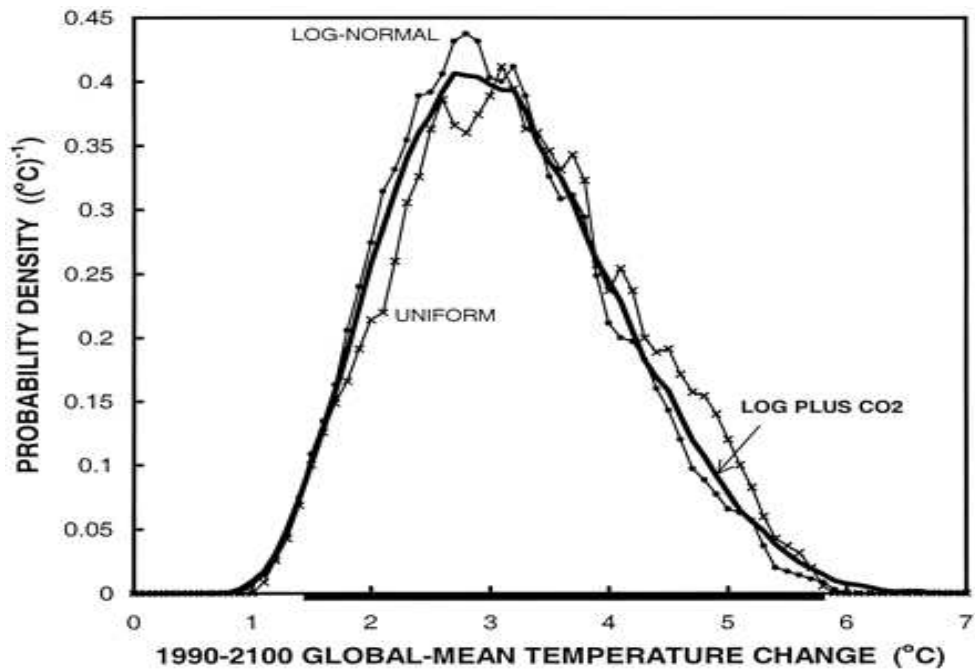

**Figure 1**  Probability distribution function of the global mean temperature increase between 1990 and 2100 (the thick line refers to the most realistic case). Taken with permission from Wigley and Raper (2001).

In Webster at al.[7] the influence on the ten-year average global mean temperature variation between 1990 and the decade 2090–2100 (Figure 2) of the uncertainties of both the natural and the economic parameters have been studied using the MIT Integrated Global System Model (IGSM; see, http://web.mit.edu/globalchange/www/if.html). This comprehends a climate model (Prinn et al., 1999) whose components are a 2-D atmosphere with explicit treatment of the chemistry of the most relevant species, a 2-D ocean, and terrestrial ecosystem. The climate model receives as inputs the greenhouse gases emissions resulting from a global economic model,[8] which is also part of the IGSM. The 90% confidence interval obtained for the 1990–2100 temperature increase is 1.1–4.5 °C, and the median value is 2.3 °C.

These studies represent a real revolution in the methodology of climate science and contain information that can be used in the implementation of policies. They represent a large leap forward with respect to the information provided in the IPCC reports.

The IPCC third report (TAR; Houghton *et al.*, 2001) for the same time frame considered in the previous studies presents for the mean global temperature increase the interval 1.4–5.8 °C without specifying the probability density function that sits over this range of variation. Such an interval has been deduced through expert elicitation based on the results obtained by the most prominent GCMs.

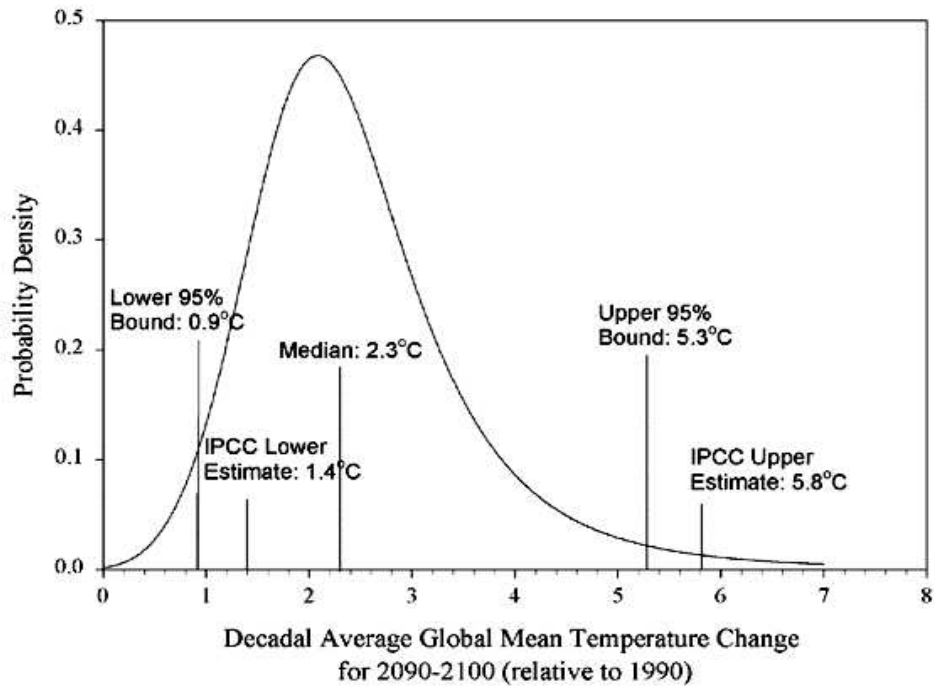

**Figure 2** Probability distribution function of the ten-year average global mean temperature increase between 1990 and the 2090–2100 decade.[7]

It is wrong to think that the previously discussed results are less relevant than those of the IPCC, because the latter have been derived from simulations of more powerful models. The previously discussed epistemology implies that only a detailed and systematic treatment of the uncertainties can provide meaningful results. Anyway, the IPCC study cannot be considered an analysis of structural uncertainties, even if the results of several models are taken into account. This is because the comparison is only made horizontally, and because no robust quantitative method of compiling the statistics for the output variable from the results of the various models has been presented. These are the reasons why a probability density function could not be proposed, thus providing the readers with a rather limited amount of information. This exemplifies that the most powerful model is not necessarily the best one to achieve any kind of goal: the flexibility, which is typical of simpler models, can be more important and effective than the ability to reproduce the details of the evolution of every single part of the system.

A new cutting-edge strategy for the study of the uncertainties in climate change has been proposed by a UK-based project led by M. Allen and D. Stainforth (see, http://www.climateprediction.net). The goal of this project, which is presently in its preliminary phase, is to explore the parametric uncertainties of various versions of the state-of-the-art 3-D climate model developed at the Hadley Meteorological Centre (see,

http://www/metoffice.gov.uk/research/hadleycentre/models/modeltypes.html). In order to overcome the problem of the huge amount of computer power needed to study the uncertainties of models presenting such an high degree of complexity, the scientists working at the www.climateprediction.net project propose to distribute the computing among thousands of ordinary PCs, belonging to private citizens or institutions that wish to participate in the experiments. The strategy is to use the idle time of the computers to perform model runs, along similar lines to the Seti@home project (see, http://www.setiathome.ssl.berkeley.edu/index.html).

The results presented by Wigley *et al*. (2001) and by Webster *et al*.[7] are extremely interesting and useful to address the task of evaluating the global costs of climate change. These are usually estimated as functions of the variations of global mean climatic variables (Nordhaus, 1994), especially of the temperature.[9] These functions contain parameters describing the ability and the possibility to adapt to and mitigate the climatic change. One of the substantially correct features of these functions is that they have a strongly non-linear structure, which can be exponential or polynomial.[10] These functional forms describe costs that almost vanish if the extent of the climatic change is small, thanks to relatively cheap mitigation and adaptation policies, while they explode if the changes exceed values beyond which an elastic response of the society is not possible anymore and the chances of bad climatic surprises is high. These functions provide a sort of quantitative framework of a precautionary principle. The fact that the costs are intrinsically non-linear with the climatic changes implies that the IPCC results do not provide sufficient information for an intelligent process of decisionmaking. Since the costs of the climate change increase by orders of magnitude, from the lower to the upper bound of the interval of variation of the global mean temperature change from 1990–2100, the detailed form of the probability distribution function of this variable is necessary to estimate the expected value of the costs. Moreover, in a sensible process of decisionmaking, in order to wisely apply the precautionary principle, it is at least necessary to know if the probability of a catastrophic scenario (e.g. $\Delta T > 5$ °C) can be reasonably estimated as being of the order of 1%, 0.01% or 0.0001%.

## 4   Conclusions

The previously presented studies do not absolutely have the last word in terms of climate change projections; their main merit is that they present an epistemologically correct instrument of investigation. In terms of research strategy, the idea of distributed computing proposed by the www.climateprediction.net research group might initiate an entirely new generation of scientific investigations on complex systems. No complete studies of the effects of model uncertainties in climate change projections have yet been done, even if very promising work in the field of quantification of uncertainties arising from the bias between atmospheric model and the actual system is ongoing (Smith, 1997; 2000).

These studies and future studies along these lines can provide useful information for the process of decisionmaking on a global scale, whose implementation is needed in the short term. It is necessary to reformulate the idea that scientific investigations can provide simple truths that eventually constitute a baseline for the implementation of policies. In any matter involving complex systems, it is necessary to accept that any political choice has to be taken in intrinsic, and thus unavoidable, conditions of uncertainty. Usually it is proposed to delay the creation of a binding legislation on greenhouse gas emissions until scientific answers of improved precision provide better-defined suggestions for policies able to prevent environmentally adverse consequences within a context of economic efficiency. This is derived from the implicit assumption that in future, in light of the improved knowledge of the climatic system, the past conservation policies would prove exceedingly stringent. This assumption that the present ideas about climate change are biased towards catastrophism is a wrongly optimistic superficial attitude. There are several evidences that in the recent and distant past the climate of our planet experienced sudden changes, and it is clear that the rapid and violent forcing due to greenhouses gas emissions enhances the chances of the manifestation of bad climatic surprises (Rahmstorf *et al.*, 2000).

## 5  Acknowledgements

The author wishes to thank N. Ashford, C. Forest, M. Hesse, G. Russell and P. Stone for useful suggestions on the form and the content of the paper, and M. Webster for permission to reproduce Figure 2.

**Endnotes**

1  'Complex' comes from the past participle of the Latin verb *complector*, *-ari* (to entwine). Note the difference between the precise meaning of 'complex' and 'complicated', which comes from the past participle of the Latin verb *complico*, *-are* (to put together).

2  The science relevant to genetically modified organisms also investigates complex systems.

3  Defined as the difference between the mean global temperature at equilibrium, corresponding to an atmospheric $CO_2$ concentration which is double and equal, respectively, to the atmospheric $CO_2$ concentration during the pre-industrial age (~280 parts per million).

4  See, for example, the paper by Hansen and Sato; available at

   http://www.giss.nasa.gov/gpol/papers/2001/2001_HansenSato.pdf.

5  On the other hand, equivalently: each being the restriction and simplification of another one, starting from an initial complex parent.

6  A given value of a parameter is used in the simulations with frequency proportional to the value of the probability distribution for such a value.

7  See Webster *et al.*; available at

   http://web.mit.edu/globalchange/www/MITJPSPGC_Rpt73.pdf.

8  M.H. Babiker *et al.*; available at

http://web.mit.edu/globalchange/www/MITJPSPGC_Rpt71.pdf.

9   The global costs should be computed as the sum of all the local costs. These essentially depend on the local features of the society, the environment and the climate change. Presently, it is not possible to give robust estimates for the latter.

10  When the globe is divided into macro regions, the value of the parameters that uniquely determine the cost functions depend on the macro region considered, in order to take into account local differences in the ability and possibility to adapt to climate change.